# MODEL OF GENERATION OF ELECTROMAGNETIC EMISSION DETECTED BEFORE EARTHQUAKE

**M.K. Kachakhidze[1], N.K. Kachakhidze[1], T.D. Kaladze[2]**

[1]*St. Andrew The First-Called Georgian University of The Patriarchy of Georgia, Tbilisi, Georgia*
[2]*Iv. Javakhishvili Tbilisi State University, Institute of Applied Mathematics, Tbilisi, Georgia*

*Correspondence to:* M. K. Kachakhidze      manana_k@hotmail.com

## Abstract

Recent satellite and ground-based observations proved that during an earthquake preparation period VLF/LF and ULF electromagnetic emissions are observed in the seismogenic area.

The present work offers possible physical bases of earth electromagnetic emission generation detected in the process of earthquake preparation. According to the authors of the present paper electromagnetic emission in radiodiapason is more universal and reliable than other earthquake indicators and VLF/LF electromagnetic emission might be declared as the main precursor of earthquake.

It is expected that in the period before earthquake namely earth electromagnetic emission offers us the possibility to resolve the problem of earthquake forecasting by definite precision and to govern coupling processes going on in lithosphere-atmosphere-ionosphere (LAI) system.

## Introduction

In the period of large earthquakes preparation, occurrence and often after their occurrence MHz, KHz, ULF electromagnetic emission is detected (Hayakawa and Fujinawa, 1994; Hayakawa, 1999; Hayakawa et al.,1999; Gershenzon and Bambakidis, 2001; Biagi, et al.,1999; . Hayakawa and Molchanov, 2002; Bahat, et al., 2005; Pulinets, et al., 2007; Eftaxia,s et al., 2007a; 2007b; 2009; Biagi, et al.,2009; Eftaxias et al., 2010; Biagi, et al., 2013).

In the present paper we'll try to explain this phenomenon on the basis of electrodynamics and create a model of generation of LAI system self-generated electromagnetic oscillation.

From the very moment of starting earthquake preparation, segment of the Earth crust where incoming earthquake focus is to be formed, belongs to the system, which suffers specific type oscillations: the process of energy accumulation is in progress in the system. But simultaneously as a result of foreshocks, main shock and aftershocks the accumulated energy is released too. With this in view this system is an oscillation system.

Extreme diversity of oscillation systems and their properties at the study of oscillation processes going on in them needs identification of common features in various oscillation systems and their grouping into certain classes and types according to the most characteristic signs.

Seismogenic zone simultaneously can be considered as a distributed system since in this system mass, elasticity (mechanical systems), capacity and induction (electric systems) elements are uniformly spread in the whole volume of the system. It should be stated that in earthquake preparation area each least element has its own capacity and induction because of piezo-electric, electrochemical and other effects.

Alongside with it, the system is considered as distributed, if time of transfer of perturbation along the system is not less than the oscillation period.

Thus, for distributed systems quasi static terms are not fulfilled in the principle. Main movements in such systems – are wave motions (Migulin V.V., et al.,1978). Due to the fact that distributed systems don't meet the demands of quasi statics, for electric field $rot\ \boldsymbol{E} \neq \boldsymbol{0}$.   In this case integral $\int_1^2 E_s ds$  depends on a path of integration between points 1 and 2, and because of it, we can't introduce the notion of a potential and capacity.

But if in double-wire or coaxial wires the term that b distance among wires is less than wire length  $l$  and wave length  $\lambda$  ($b \ll l, b \ll \lambda$), is satisfied, in case of low resistance of conductors, there will be only transverse electromagnetic waves (Migulin V.V., et al.,1978).

Thus, in the plane, normal to the line, distribution of these waves will coincide with distribution of electric and magnetic fields for static case. Therefore for small sections of $dx$ line we can consider theory of quasi static currents as acceptable and can introduce a notion of a potential, current, distributed capacity and induction.

Since the processes going on in earthquake focal zone in the earthquake preparation period satisfy the demands of distributed systems, there should be the linear  induction  $L_0$ and capacity  $C_0$ there. Thus, processes taking place in the period of earthquake preparation coincide, by definite precision, with the above described distributed system for double-wire lines.

Values of  linear induction $L_0$ and capacity  $C_0$ (that is induction and capacity per length unit) are determined  by geometry of conductors and properties of  medium.  For double-wire lines, if a distance between the wire is $b$, electromagnetic wave length  $\lambda$  and  if the term $b \gg r$  is satisfied where $r$  is a wire radius, for double-wire system inductivity we'll have (Migulin V.V., et al.,1978):

$$L_0 = \frac{4\mu}{c^2} \ln \frac{b}{r} \qquad (1)$$

where $c$ is velocity of  light in vacuum,  $\mu$  medium magnetic permeability, and for capacity:

$$C_0 = \frac{\varepsilon}{4\ln\left(\frac{b}{r}\right)} \qquad (2)$$

where  $\varepsilon$ is medium dielectric permeability.

## Description of a model

It has been proved experimentally that at the formation of cracks in the period of earthquake preparation electric dipoles appear on their surface (Freund, et al. 2006; Eftaxias, et al., 2007a; 2007 b).

In solid medium accumulation of significant quantity of polarization charge  may take place in areas, where heterogenity of definite linear scale are formed already or are in the process of formation.

Polarization effect is often accompanied by electromagnetic emission (Ikeya and Takaki, 1996; Yoshida et al., 1997) and this, formally is a sign, that  besides electrostatic effect, which forms and creates capacity, the polarization is accompanied by induction effect too. But at the analysis of a possibility of induction  interaction  in  lithosphere-atmosphere  system  we  have  to  take  into consideration that there are many other possibilities of development of  induction effect. Probably, with respect to seismic phenomena, the source of this effect is always lithosphere. Therefore, we can assume that schematically we have to deal with definite type electromagnetic contour, elements of which should be connected with lithosphere as well as with atmosphere.

Conditionally it is admitted that the earth surface has negative potential to atmosphere; therefore , before piezo-effect, which is conditioned in rocks by mechanical tensions (Mognaschi, 2002; Triantis et al., 2008;Telesca et al., 2013), the segment of lithosphere, where earthquake prepares, can be considered as negatively charged.

According to the avalanche-like unstable model (Mjachkin, et al., 1975), in a focal zone of incoming earthquake, on the second stage of its preparation, at the background of multiple cracking, nucleus of definite linear size of main fault is beginning to originate, length of which is gradually increasing. Therefore this main fault nucleus can be imagined as a conductor, length of which exceeds significantly the characteristic size of its cross section.

As a result of growth of tectonic stress the heterogenity, that is, zones of positive charge appear in earthquake preparation areas (Bleier, et al. 2009). Similar to "Frankel's generator" in this segment of the Earth crust we'll have induction polarization (Yoshino, 1991; Molchanov et al., 1993; Hayakawa et al., 2002, Liperovsky et al., 2008).

Generally, polarization charge should be distributed over some surface, which should be limited by fault or should be formed along the faults (Yoshino, 1991), that is, near the first conductor the same size polarized conductor should formed by induction.

Due to the fact that earthquake preparation takes place in a relatively weak zone, with the view of solidity (Mjachkin, et al., 1975; Morozova, et al., 1999; Tada-nori Goto, et al., 2005; Kovtun, et al., 2009), segment of earthquake preparation should be surrounded from top to bottom by rocks of relatively high solidity. Although, considering that generally rock density at the increase of depth is growing, rocks existing above the zone of future main fault will have lower density, than the lower rocks.

In such case, rate of heterogenity because of impact of one and the same tectonic stress will be significantly high in less dense, that is, in upper rock. Since accumulation of polarization charge is connected with heterogenities existing in rocks, the second polarization conductor should be formed above the stated fault, quasi-parallel to it. Thus the main fault can be imagined as double-wire conductor, the length of which exceeds greatly the size of cross section characteristic to it.

There are data in special scientific literature which states that during earthquake preparation period, in focal zone of an earthquake there are galvanic effects that can be regarded as seismogalvanic effects (Moroz et al., 2001). It should be taken into consideration that in focal zone there are thermally anomalous sections too.

Alongside with it observations show change of specific electric resistance of rocks, which have the character of earthquake precursors (Sumitomo and Noritomi, 1986, Wang Zhiliang, Yu Surong, 1989; Du Xuebin, 1992; Bragin, et al., 1992);

According to recent studies at the depth of 5-10 km, in seismic faults plane and in its vicinity conductivity increases significantly, because, irrespective of the fact that the medium is heterogeneous, there are inclusions of high electric conductivity in the form of graphitization and sulphurizing of rocks (Morozova, et al., 1999; Tada-nori Goto, et al., 2005; Kovtun, et al., 2009). At such conditions the above referred double-wire conduction layer can be closed in the Earth crust, in high conductivity zones (layers) and enable virtual conductors to create a contour.

And really, if formally there are two, separated from each other, horizontal wires of opposite polarity in such heterogeneous medium, where there are high electric conduction inclusions in the form of graphitization and sulphurizing, a structure similar to vibrational contour should be formed, which might be locked by vertical electric field.

The fact should be emphasized that at the second stage of earthquake preparation, when avalanche –like cracking process takes place, tectonic stress value in the earthquake focus is very close to the threshold level, which is obligatory to overcome geological strength limit of medium. Irrespective of constant growth of tectonic stress, earthquake doesn't occur yet, since definite share of tectonic energy is used for avalanche formation of cracks and their aggregation. At this stage, resource of mechanical energy can be considered constant, by definite precision. At such terms, the above stated distributed system will act in parallel, as conservative system, which, in its turn is an idealized system and is characterized by constant resource of mechanical or electromagnetic energies, or jointly, both of these energies in the process of oscillations. Since one of the simplest form of a conservative system is pendulum and electric oscillation contour, and earthquake focal zone at the final stage of earthquake preparation combines, at a definite accuracy, the properties of namely this system, it might be allowed to characterize in a similar way mechanical and electric oscillation processes going on in it.

Therefore, if we don't take into consideration electromagnetic dissipation, we can use the known formula for determination of contour's self-generated oscillation frequency:

$$\omega^2 = \frac{1}{L \cdot C} \qquad (3)$$

Since at this stage of earthquake preparation earthquake focus already simultaneously combines properties of distributed and conservative systems, capacity and inductivity of connecting wires will play significant role and reciprocal capacity and inductivity of connecting wires (linear conductors) will be quite sufficient for generation of electromagnetic oscillations. At the same time, it is not obligatory for virtual wires be strictly closed in contour frames. The main term is existence of wire closing mechanism.

Of course irrespective of disregarding of ohm resistance effect in contour, there will necessarily be energy loss because of electromagnetic emission, the intensity and direction of spreading of which will be depended on contour form and spatial dimensions.

Assuming that a distance between wires is $b$ and $b \gg r$, where $r$ is a wire radius, $L$ and $C$ are induction and capacity of $l$ length double wire correspondingly, taking (1) and (2) into account, the formula (3) will be written as follows:

$$\omega = \frac{1}{\sqrt{\varepsilon\mu}} \times \frac{c}{l} = k\frac{c}{l} \quad (4)$$

where $c$ is velocity of light and $k = \frac{1}{\sqrt{\varepsilon\mu}}$.

Due to the fact that measurements are made in earth surface boundary layer and for air $\varepsilon = \mu = 1$ (the SI system is used), will be

$$\omega = \frac{c}{l} \quad (5)$$

Let's allow that $l$ changes within the interval (1-300) $km$, that corresponds to the diapason of alteration of size, characteristic to fault in earthquake focus. From (5) the diapason of the self-generated electromagnetic oscillation frequency changes of the analogous contour will be $\omega = (10^3 - 10^5)$ Hz, which is in good quantitative conformity with the spectrum of electromagnetic emission frequencies which are often fixed in the period before earthquake (Boudjada, et.al., 2010).

**Discussion and conclusion**

The present work offers interpretation of a mechanism of formation of hypothetic ideal electromagnetic contour, creation of which is envisaged in incoming earthquake focal zone. Model of generation of EM emission detected before earthquake is based on physical analogues of distributed and conservative systems and focal zones. According to the model the process of earthquake preparation from the moment of appearance of cracks in the system, including completion of series of foreshocks, earthquake and aftershocks, are entirely explained by oscillating systems.

At the starting stage of earthquake preparation focal zone is both a mechanical distributed system and electric distributed system. At the final stage of earthquake preparation focal zone combines the properties of conservative system too. Energy release in the system is performed by the characteristic to conservative system signs: by mechanical vibrations and electromagnetic emission. It is considerable that as soon as the earthquake occurs and mechanical energy is emitted, focal zone loses properties of mechanical conservative system. When earthquake occurs, focal zone loses the properties of electric conservative system too, since contour is destroyed, and correspondingly, no electromagnetic emission takes place.

The paper offers a formula of dependence of the main faults length from the frequency of emitted electromagnetic wave. The offered model gives qualitative explanation of a mechanism of generation of electromagnetic waves emitted in the earthquake preparation period.

**References:**

Bahat, D., Rabinovitch, A., and Frid, V.: Tensile fracturing in rocks: Tectonofractographic and Electromagnetic Radiation Methods, Springer Verlag, Berlin, 570 pp., 2005;

Biagi, P.F., Seismic Effects on LF Radiowaves, In: Atmospheric and Ionospheric Electromagnetic Phenomena Associated with Earthquakes, (Ed.: M. Hayakawa), TERRAPUB, Tokyo, 535-542, 1999;

Biagi, P. F., Castellana, L., Maggipinto, T., Loiacono, D., Schiavulli, L., Ligonzo, T., Fiore, M., Suciu, E., and Ermini, A.: A pre seismic radio anomaly revealed in the area where the Abruzzo earthquake (M = 6.3) occurred on 6 April 2009, Nat. Hazards Earth Syst. Sci., 9, 1551–1556, 2009. doi:10.5194/nhess-9- 1551-2009;

Biagi, P.F., Magippinto, T., Schiavulli L., Ligonzo,. T. Ermini, A. European Network for collecting VLF/LF radio signals (D5.1a). DPC- INGV -S3 Project. "Short Term Earthquake prediction and preparation", 2013;

Bleier, T., Dunson, C., Maniscalco, M., Bryant, N., Bambery, R., and Freund, F.: Investigation of ULF magnetic pulsations, air conductivity changes, and infra red signatures associated with the 30 October Alum Rock M5.4 earthquake, Nat. Hazards Earth Syst. Sci., 9, 585–603, 2009. doi:10.5194/nhess-9-585-2009;

Bragin, V.D., Volykhin, A.M., Trapeznikov, Yu.A. Electrical resistivity variations and moderate earthquakes, Tectonophysics, 202 (2-4) 233-238, 1992;

Boudjada, M.Y.; Schwingenschuh,K., Döller, R., Rohznoi, A., Parrot,M., Biagi, P. F., Galopeau, P.H.M., Solovieva, M., Molchanov,O., Biernat,H.K., Stangl, G, Lammer, H., Moldovan, I. Voller.W, and Ampferer, M. Decrease of VLF transmitter signal and Chorus-whistler waves before l'Aquila earthquake occurrence. Nat. Hazards Earth Syst. Sci., 10, 1487–1494, 2010. www.nat-hazards-earth-syst-sci.net/10/1487/2010/. doi:10.5194/nhess-10-1487-2010;

Du Xuebin. On sudden change of the Earth resistivity preceding the moderately strong or strong earthquake in China continent and its temporal- spatial distribution - a series of studies on the short-term and impending earthquake prediction, Earthquake (Beijing), 6 (6) 51-6, 1992;

Eftaxias, K., Sgrigna, V., and Chelidze, T.: Mechanical and Electromagnetic Phenomena Accompanying Preseismic Deformation: from Laboratory to Geophysical Scale, Tectonophysics, 431, 1– 301, 2007a;

Eftaxias, K., Panin, V., and Deryugin, Y.: Evolution EM-signals before earthquake and during laboratory test of rocks, Tectonophysics, 431, 273–300, 2007b;

Eftaxias, K., Athanasopoulou, L., Balasis, G., Kalimeri, M., Nikolopoulos, S., Contoyiannis, Y., Kopanas, J., Antonopoulos, G., and Nomicos, C.: Unfolding the procedure of characterizing recorded ultra low frequency, kHZ and MHz electromagnetic anomalies prior to the L'Aquila earthquake as preseismic ones – Part 1, Nat. Hazards Earth Syst. Sci., 9, 1953–1971,2009. doi:10.5194/nhess-9-1953-2009;

Eftaxias, K., Balasis, G., Contoyiannis, Y., Papadimitriou, C., Kalimeri, M., Athanasopoulou, L., Nikolopoulos, S., Kopanas, J., Antonopoulos, G., and Nomicos, C.: Unfolding the procedure of characterizing recorded ultra low frequency, kHZ and MHz electromagnetic anomalies prior to the L'Aquila earthquake as pre-seismic ones – Part 2, Nat. Hazards Earth Syst. Sci., 10, 275–294,2010. doi:10.5194/nhess-10-275-2010;

Freund, F. T., Takeuchi, A., and Lau, B. W. S.: Electric currents streaming out of stressed igneous rocks – A step towards understanding pre-earthquake low frequency EM emissions, Phys. Chem. Earth, 31, 389–396, 2006;

Gershenzon, N. and Bambakidis, G.: Modelling of seismoelectromagnetic phenomena, Russian Journal of Earth Sciences, 3, 247–275, 2001;

Hayakawa, M. and Fujinawa, Y.: Electromagnetic Phenomena Related to Earthquake Prediction, Terrapub, Tokyo, 1994;

Hayakawa, M.: Atmospheric and Ionospheric Electromagnetic Phenomena Associated with Earthquakes, Terrapub, Tokyo, 1999;

Hayakawa, M., Itoh, T., and Smirnova, N.: Fractal analysis of ULF geomagnetic data associated with the Guam earthquake on August 8, 1993, Geophys. Res. Lett., 26, 2797–2800, 1999;

Hayakawa, M. and Molchanov, O. A.: Seismo-Electromagnetics: Lithosphere-Atmosphere-Ionosphere Coupling, TERRAPUB, Tokyo, 1–477, 2002;

Ikeya, M., Takaki, S.. Electromagnetic fault for earthquake lightning, Japan Journal of Applied Physics Part 2:Letters, 35 (3A) L355-L357, 1996;

Kovtun A.A. Principles of the Earth Physics. Electrical conductivity of the Earth. Educational-methodological handbook for advanced training course students in specialty "Geophysics" according to the program "Methods of search and prospecting of deposits of useful minerals in industrial and prospecting geophysics " Kazan State University, Kazan, pp. 51, 2009 (In Russian);

Liperovsky, V. A., Meister, C.-V., Liperovskaya, E. V., and Bogdanov, V. V.: On the generation of electric field and infrared radiation in aerosol clouds due to radon emanation in the atmosphere before earthquakes, Nat. Hazards Earth Syst. Sci., 8, 1199–1205,2009. doi:10.5194/nhess-8-1199-2008, 2008;

Migulin V.V., Medvedev V.I., Mustel E.P., Parigin V.N.Principles of oscillation theory. Moscow, "Nauka". Central editorial staff for physical-mathematical literature. 391 p., 1978 (In Russian). Mjachkin, V.I., Brace W.F., Sobolev G.A., Dieterich J.H. Two models for earthquake forerunners, Pageoph., vol. 113. Basel, 1975;

Mognaschi, E. R.: IW2GOO. On The Possible Origin, Propagation And Detectebility Of Electromagnetic Precursors of Eaerthquakes, Atti Ticinensi di Scienze della Terra, 43, 111–118, 2002;

Molchanov, O. A., Mazhaeva, O. A., Golyavin, A. N., and Hayakawa, M.: Observation by the Intercosmos-24 Satellite of ELF-VLF electromagnetic emissions associated with earthquakes, Ann. Geophys., Atmos. Hydrospheres Space Sci., 11, 5,431–440, 1993;

Moroz, Yu.F., Vershinin, E.F. Magnetotelluric monitoring of the seismically active Kamchatka region, Izvestiya - Physics of the Solid Earth, 37 (1) 832-838, 2001;

Morozova G.M., Menstein A.K. Eltsov I.N. Depth electromagnetic probing with controlled source in Baikal rift zone // Geophysical methods of study of the Earth crust" Coll. scientific reports of All-Russia Geophysical Conference. RAS. Siberian Dept. United Institute of Geology, Geophysics and Mineralogy/ Ed.: S.V. Goldin, A.D. Duchkov, Novosibirsk, p.57-62, 1999 (In Russian);

Pulinets, S. A., Biagi, P., Tramutoli, V., Legen'ka, A. D., and Depuev, V. K.: Irpinia earthquake 23 November 1980 – Lesson from Nature reviled by joint data analysis, Ann. Geophys., 50(1), 61–78, 2007;

Sumitomo, N., Noritomi, K. Synchronous precursors in the electrical earth resistivity and the geomagnetic field in relation to an earthquake near the Yamasaki fault, Southwest Japan., Journal of Geomagnetism & Geoelectricity, 38 (1) 971-989, 1986;

Tada-nori Goto, Yasuo Wada, Naoto Oshiman, Norihiko Sumitomo, Resistivity structure of a seismic gap along the Atotsugawa Fault, Japan. Physics of the Earth and Planetary Interiors, 148 55–72, ELSEVIER, 2005;

Telesca L., Lapenna V., Balasco M., Rpmano G., Siniscalchi A. Review of the main results and state of the art optimal experimental methodologies for the study of electromagnetic field variations associated to seismogenic processes. DPC-INGV-S3 Project. ''Short Term Earthquake prediction and preparation''. INGV, 2013;

Triantis, D., Anastasiadis, C., and Stavrakas, I.: The correlation of electrical charge with strain on stressed rock samples, Nat. Hazards Earth Syst. Sci., 8, 1243–1248, 2008. doi:10.5194/nhess-8-1243- 2008;

Wang Zhiliang, Yu Surong. Characteristics of earth resistivity anomalies at Tengchong and other stations before the Lancang-Gengma M 7.6 earthquake, Earthquake (Beijing), (2) 42-47, 1989;

Yoshida, S., Uyeshima, M., Nakatani, M. Electric potential changes associated with slip failure of granite: Preseismic and coseismic signals, Journal of Geophysical Research B: Solid Earth, 102 (B7) 14883-14897, 1997;

Yoshino, T.: Low-Frequency Seismogenic Electromagnetic Emissions as Precursors to Earthquakes and Volcanic Eruptions in Japan, Journal of Scientific Exploration, 5(I), 121–144, 1991.